\documentstyle[twocolumn,prl,aps,epsf]{revtex}

\newcommand{\be}{\begin{equation}} 
\newcommand{\ee}{\end{equation}}
\newcommand{\ba}{\begin{eqnarray}}
\newcommand{\ea}{\end{eqnarray}}

\title{Large $N$ dynamics in QED in a magnetic field}

\author{V.P.~Gusynin}
\address{Bogolyubov Institute for Theoretical Physics,
        03143, Kiev, Ukraine}

\author{V.A.~Miransky$^{*}$}

\address{Department of Applied Mathematics, University of Western
Ontario, London, Ontario N6A 5B7, Canada}

\author{I.A.~Shovkovy$^{*}$}

\address{Institut f\"ur Theoretische Physik,
Johann Wolfgang Goethe--Universit\"at,
60054 Frankfurt/Main, Germany}

\date{\today}

\draft
\begin{document}
\maketitle

\begin{abstract} 

The expression for the dynamical mass of fermions in QED in a magnetic
field is obtained for a large number of the fermion flavor $N$ in the
framework of $1/N$ expansion.  The existence of a threshold value
$N_{thr}$, dividing the theories with essentially different dynamics, is
established. For the number of flavors $N \ll N_{thr}$, the dynamical mass
is very sensitive to the value of the coupling constant $\alpha_b$,
related to the magnetic scale $\mu = |eB|$. For $N$ of order $N_{thr}$ or
larger, a dynamics similar to that in the Nambu-Jona-Lasinio model with
cutoff of order $|eB|$ and the dimensional coupling constant $G \sim
1/(N|eB|)$ takes place. In this case, the value of the dynamical mass is
essentially $\alpha_b$ independent (the dynamics with an infrared stable
fixed point). The value of $N_{thr}$ separates a weak coupling dynamics
(with $\tilde{\alpha}_b \equiv N\alpha_b \ll 1$) from a strong coupling
one (with $\tilde{\alpha}_b \gtrsim 1$) and is of order $1/\alpha_b$.

\end{abstract}


\section{Introduction}

The phenomenon of the magnetic catalysis of dynamical symmetry breaking
was established as a universal phenomenon in a wide class of $(2+1)$- and
$(3+1)$-dimensional relativistic models in Refs.~\cite{GMS1,GMS2} (for
earlier consideration of dynamical symmetry breaking in a magnetic field
see Refs.~\cite{Kawati,Klim}). The general result states that a constant
magnetic field $B$ leads to the generation of a fermion dynamical mass (a
gap in a one-particle energy spectrum) even at the weakest attractive
interaction between fermions. The essence of this effect is the
dimensional reduction $D\to D-2$ in the dynamics of fermion pairing in a
magnetic field. At weak coupling, this dynamics is dominated by the lowest
Landau level (LLL) which is essentially $(D-2)$-dimensional
\cite{GMS1,GMS2}. The applications of this effect have been considered
both in condensed matter physics \cite{thermo,Khvesh} and cosmology (for
reviews see Ref.~\cite{reviews}).

The phenomenon of the magnetic catalysis was studied in gauge theories, in
particular, in QED \cite{QED1,QED2,Ng,Hong,Ferrer,GS,AFK} and in QCD
\cite{Sh,Eb,KLW,MS}. In Ref.~\cite{QED2}, the present authors derived an
asymptotic expression for the fermion dynamical mass in the chiral limit
in QED, reliable for a weak coupling $\alpha_{b}$ and for the number of
charged fermions $N$ being not too large (here $\alpha_{b}$ is the running
coupling related to the magnetic scale $\mu^2 \sim |eB|)$.  Specifically,
when the parameter $\tilde{\alpha}_{b} \equiv N\alpha_{b}$ is small, i.e.,
$\tilde{\alpha}_{b}\ll 1$, the fermion dynamical mass is \cite{QED2}
\begin{equation}
m_{dyn} = C_{1}\sqrt{|eB|}
F(\tilde{\alpha}_{b})\exp\left[-\frac{\pi N}
{\tilde{\alpha}_{b}\ln\left(C_2/\tilde{\alpha}_{b}\right)}\right], 
\label{mass}
\end{equation} 
where $F(\tilde{\alpha}_b) \simeq (\tilde{\alpha}_b)^{1/3}$, and the
constants $C_1$ and $C_2$ are of order one.

In this paper, we will extend the analysis of Ref.~\cite{QED2} to the case
with a large coupling $\tilde{\alpha}_{b}$. As it will be discussed below,
such a strong coupling regime can be put under control for large
values of $N$ in the framework of $1/N$ expansion. It will be shown that
the expression for the dynamical mass in this dynamical regime is
essentially different from that in equation (\ref{mass}) and it reads:
\begin{equation}
m_{dyn} \simeq \sqrt{|eB|} \exp\left(-N\right).
\label{massN}
\end{equation}
It is noticeable that this expression of $m_{dyn}$ is $\alpha_b$
independent. As it will be shown below, the origin of such a dramatic
change of the form of the dynamical mass is intimately connected with the
dynamics of screening of the photon interactions in a magnetic field in
the region of momenta relevant for the chiral symmetry breaking dynamics,
$m_{dyn}^2 \ll |k^2| \ll |eB|$. In this region, photons acquire a mass
$M_{\gamma}$ of order $\sqrt{N\alpha_{b}|eB|}$. More rigorously,
$M_{\gamma}$ is the mass of a fermion-antifermion composite state coupled
to the photon field. The appearance of such mass resembles pseudo-Higgs
effect in the $(1+1)$-dimensional {\it massive} QED (massive Schwinger
model) \cite{Sch} (see below). The crossover from the dynamics
corresponding to expression (\ref{mass}) to that corresponding to
expression (\ref{massN}) occurs for such a threshold value of $N_{thr}$
when the mass $M_{\gamma} \sim \sqrt{N_{thr}\alpha_{b}|eB|}$ becomes of
order $\sqrt{|eB|}$, i.e., for $\tilde{\alpha}_{b}^{thr} \equiv
N_{thr}\alpha_{b} \sim 1$.

Let us consider this point in more detail. There are generically three
different scales, $\sqrt{|eB|}$, $M_{\gamma}$, and $m_{dyn}$, in this
problem. These scales correspond to the following four, dynamically
different, energy regions. The first one is the region with the energy
scale above the magnetic scale $\sqrt{|eB|}$. In that region, the dynamics
is essentially the same as in QED without a magnetic field. In particular,
the running coupling increases logarithmically with increasing the energy
scale there. The second region is that with the energy scale below the
magnetic scale $\sqrt{|eB|}$ but larger than the photon mass $M_{\gamma}$.
In that region the photon can be considered as approximately massless. The
next, third, region is the region with the energy scale less than the
photon mass $M_{\gamma}$ but larger than the fermion mass $m_{dyn}$.  In
this region, the photon is heavy, and the interaction is similar to that
in the Nambu-Jona-Lasinio (NJL) model 
(with the current-current interaction)
in a magnetic field. The important
point is that just those third and second regions are relevant for
spontaneous chiral symmetry breaking in this problem. At last, the fourth
region is the region with the energy scale $E$ less than the fermion mass
$m_{dyn}$. In that region, fermions decouple and their interaction is
suppressed by powers of the ratio $E/m_{dyn}$.

Now, when $N$ grows up to $N_{thr} \sim 1/\alpha_{b}$, the photon mass
$M_{\gamma}$ becomes of the order of the scale $\sqrt{|eB|}$ and,
therefore, the third region, between $M_{\gamma}$ and $\sqrt{|eB|}$,
shrinks and disappears. Thus for $N$ of order $N_{thr}$ or larger, the
dynamics of spontaneous chiral symmetry breaking is solely provided in the
second region, through the interaction with a heavy photon. As a result,
the dynamics becomes similar to that in the NJL model in a magnetic field
with cutoff $\sqrt{|eB|}$ and the dimensional coupling constant $G \simeq
\alpha_{b}/M_{\gamma}^2 \sim 1/N|eB|$. This implies that this dynamics is
$\alpha_b$ independent and, therefore, corresponds to an infrared stable
fixed point. It also explains the origin of the threshold value $N_{thr}
\sim 1/\alpha_{b}$.  In the rest of the paper, we will derive expression
(\ref{massN}) and justify this qualitative dynamical picture.

\section{Magnetic catalysis in QED}

We begin by considering the Schwinger-Dyson (gap) equation for 
the fermion propagator. It has the following form:
\ba
G^{-1}(x,y) &=& S^{-1}(x,y) + 4\pi\alpha_{b} \gamma^{\mu}
\nonumber\\
&\times& \int G(x,z) \Gamma^{\nu}(z,y,z^{\prime}) 
{\cal D}_{\nu\mu}(z^{\prime},x) 
d^{4} z d^{4} z^{\prime}, 
\label{SD}     
\ea
where $S(x,y)$ and $G(x,y)$ are the bare and full fermion propagators
in an external magnetic field, ${\cal D}_{\nu\mu}(x,y) $ is the 
full photon propagator and $\Gamma^{\nu}(x,y,z)$ is the full amputated
vertex function. 

Let us first consider the weak coupling dynamics ($\alpha_{b}\ll 1$) with
the number of fermion flavors $N$ of order one.  In this case, one might
think that the rainbow (ladder) approximation is reliable in this problem.
However, this is not the case. Because of the $(1+1)$-dimensional form of
the fermion propagator in the LLL approximation, there are relevant higher
order contributions \cite{QED1,QED2}. In particular, there is a large
contribution of fermions to the polarization operator. Fortunately, one
can solve this problem \cite{QED2}. Let us discuss this in more detail.

First of all, one can show that the dynamics of the fermion-antifermion
pairing is mainly induced in the region of momenta $k$ much less than
$\sqrt{|eB|}$ and much larger than the dynamical mass $m_{dyn}$, i.e., in
the the second and third scale regions discussed in Introduction. In
particular, this implies that the magnetic scale $|eB|$ yields a dynamical
ultraviolet cutoff in this problem.

The important ingredient of this dynamics is a large contribution of
fermions to the polarization operator. It is large because of an
(essentially) $(1+1)$-dimensional form of the fermion propagator in a strong
magnetic field. Its explicit form in the one-loop approximation is
\cite{QED2}:
\ba
{\cal P}^{\mu\nu} \simeq \frac{\alpha_{b}N}{3\pi}
\left(k_{\parallel}^{\mu}
k_{\parallel}^{\nu}-k_{\parallel}^{2}g_{\parallel}^{\mu\nu}\right)
\frac{|eB|}{m^{2}_{dyn}},
\label{Pi-IR}
\ea
for $|k_{\parallel}^2| \ll m_{dyn}^2$, and
\ba
{\cal P}^{\mu\nu} \simeq -\frac{2\alpha_{b}N}{\pi}
\left(k_{\parallel}^{\mu}
k_{\parallel}^{\nu}-k_{\parallel}^{2}g_{\parallel}^{\mu\nu}\right)
\frac{|eB|}{{k_{\parallel}^2}}, 
\label{Pi-UV}
\ea
for $m_{dyn}^2 \ll |k_{\parallel}^2|\ll |eB|$, 
where $g_{\parallel}^{\mu\nu}\equiv \mbox{diag}(1,0,0,-1)$ is the
projector onto the longitudinal subspace, and $k_{\parallel}^{\mu}\equiv
g_{\parallel}^{\mu\nu} k_{\nu}$ (note that the magnetic field is in the
$x^3$ direction). Similarly, we introduce the orthogonal projector
$g_{\perp}^{\mu\nu}\equiv g^{\mu\nu} -g_{\parallel}^{\mu\nu}
=\mbox{diag}(0,-1,-1,0)$ and $k_{\perp}^{\mu}\equiv g_{\perp}^{\mu\nu}
k_{\nu}$ that we shall use below. Notice that fermions in a strong
magnetic field do not couple to the transverse subspace spanned by
$g_{\perp}^{\mu\nu}$ and $k_{\perp}^{\mu}$. This is because in a strong
magnetic field only the fermions from the LLL matter and they couple only
to the longitudinal components of the photon field. The latter property
follows from the fact that spins of the LLL fermions are polarized along
the magnetic field \cite{QED1}.

The expressions (\ref{Pi-IR}) and (\ref{Pi-UV}) coincide with those for
the polarization operator in the massive $QED_{1+1}$ (Schwinger model)
\cite{Sch} if the parameter $2\alpha_{b}|eB|$ here is replaced by the
dimensional coupling $e^{2}_{1}$ of $QED_{1+1}$. As in the Schwinger
model, Eq.~(\ref{Pi-UV}) implies that there is a massive resonance in the
$k_{\parallel}^{\mu}k_{\parallel}^{\nu} -k_{\parallel}^{2}
g_{\parallel}^{\mu\nu}$ component of the photon propagator. 
Its mass is
\be
M_{\gamma}^2= \frac{2 N\alpha_{b}}{\pi}|eB|.
\label{M_g}
\ee
This is reminiscent of the pseudo-Higgs effect in the 
$(1+1)$-dimensional massive QED. It is not the genuine Higgs effect 
because there is no complete screening of the electric charge in the 
infrared region with $|k_{\parallel}^2|\ll m_{dyn}^2$. This can 
be seen clearly from Eq.~(\ref{Pi-IR}). Nevertheless, the pseudo-Higgs 
effect is manifested in creating a massive resonance and this 
resonance provides the dominant forces leading to chiral
symmetry breaking.  

Now, the main points of the analysis of the weak coupling dynamics in QED
in a magnetic field are \cite{QED2}: (i) the so called improved rainbow
approximation is reliable in this problem provided a special non-local
gauge is used, and (ii) the relevant region of momenta in this problem is
$m_{dyn}^2 \ll |k^2| \ll |eB|$. We recall that in the improved rainbow
approximation the vertex $\Gamma^{\nu}(x,y,z)$ is taken to be bare and the
photon propagator is taken in the one-loop approximation. For a weak
coupling dynamics, this approximation is reliable since in that special
gauge the loop contributions in the vertex are suppressed by powers of
$\alpha_b$.  [It is appropriate to call this approximation the
``strong-magnetic-field-loop improved rainbow approximation". It is an
analog of the hard-dense-loop improved rainbow approximation in QED or QCD
with a nonzero baryon density \cite{hard}]. This leads us to the
expression (\ref{mass}) for the dynamical gap.

Let us now turn to the case with a large number of fermion flavors $N$.
The crucial point is that the improved rainbow approximation is still
reliable in this case. The essential difference, however, is that now one
has to consider not the conventional loop expansion (with a small
$\alpha_b$) but the $1/N$ expansion (with a small $1/N$). It is well known
\cite{Coleman} that in this expansion the coupling constant
$\tilde{\alpha}_{b}\equiv N\alpha_{b}$ has to be kept fixed as $N \to
\infty$. A great advantage of the $1/N$ expansion is that now one can
treat the dynamics with an arbitrary value of $\tilde{\alpha}_{b}$: it
could be small $(\tilde{\alpha}_{b} \ll 1)$, intermediate
$(\tilde{\alpha}_{b} \sim 1)$, or large $(\tilde{\alpha}_{b} \gg 1)$.
Indeed, independently of the value of $\tilde{\alpha}_{b}$, the loop
corrections in the vertex are suppressed by powers of $1/N$ and,
therefore, the improved rainbow approximation is indeed reliable for 
large $N$.

Let us now proceed to the analysis of the SD equation for the dynamical
mass of fermions in QED in a magnetic field for a large number of flavors
$N$. In the improved rainbow approximation, the SD equation reads in
Euclidean space [see Eq.~(54) in Ref.~\cite{QED2}]:
\begin{equation}
B(p^2)=\frac{\alpha_{b}}{2\pi^2}
\int\frac{d^2q B(q^2)} {q^2+m_{dyn}^2}
\int\limits_{0}^{\infty}\frac{dx \exp(-x/2|eB|)}
{x+({\bf q}-{\bf p})^2+M_\gamma^2},
\label{inteqlinear}
\end{equation}
where $B(q^2)$ is the fermion mass function and the two-dimensional 
vector ${\bf q}$ is ${\bf q}=(q_4,q_3)$ with $q_4=-iq_0$.

As we mentioned in Introduction, in the limit of small coupling constant
$\tilde{\alpha}_b$, the above SD equation was solved in Ref.~\cite{QED2},
using numerical as well as approximate analytical methods. The result for
the dynamical mass of fermions is quoted in Eq.~(\ref{mass}). Here we
would like to comment on the nature of the interaction, provided by
photons, in this weak coupling regime. One could easily check that the 
dominant interaction is provided by the photons with the (``longitudinal")  
momenta in the following range: $m_{dyn}^2 \alt ({\bf q}-{\bf p})^2 \alt
|eB|$. Then, by noticing that the photon mass also lies in the same range
of momenta, i.e., $m_{dyn}^2 \ll M_{\gamma}^2 \ll |eB|$, one finds that
the degree of importance of the photon mass is changing when the values of
momenta are sweeping the relevant range of momenta. While in the
near-infrared region with $m_{dyn}^2 \alt ({\bf q}-{\bf p})^2 \alt
M_{\gamma}^2$ (the third region, in the nomenclature of Introduction) the
interaction is local with a good precision, it becomes essentially
nonlocal in the intermediate range of momenta where $M_{\gamma}^2 \alt
({\bf q}-{\bf p})^2 \alt |eB|$ (the second region in that nomenclature).

In the opposite limit, $\tilde{\alpha}_b \gtrsim 1$, the structure of 
the SD equation (\ref{inteqlinear}) considerably simplifies. The
simplification comes due to the new hierarchy of scales, $|eB| \lesssim
M_{\gamma}^2$ (see Eq. (\ref{M_g})). From physical point of view, 
this hierarchy means that the photon mass 
is so large that the interaction leading to fermion pairing is
essentially local. Therefore, by neglecting $({\bf q}-{\bf p})^2$
term in the denominator of the second integral on the right hand side of
Eq.~(\ref{inteqlinear}), we derive an approximate algebraic form of the
gap equation that works rather well at large values of
$\tilde{\alpha}_b$,
\be
-\frac{\tilde{\alpha}_b}{2 \pi N}
\exp\left(\frac{\tilde{\alpha}_b}{\pi}\right)
\mbox{Ei}\left(-\frac{\tilde{\alpha}_b}{\pi}\right)
\ln\frac{|eB|}{m^2} =1,
\label{gap-eq-N}
\ee
where $\mbox{Ei}(z)$ is the exponential integral function. By making use
of the asymptotic expansion of the exponential integral function at large
$\tilde{\alpha}_b$, this equation is further simplified, and the
following result for the dynamical mass of fermions is obtained:
\be
m_{dyn} \simeq \sqrt{|eB|} \exp\left(-N\right), \quad
\mbox{for} \quad \tilde{\alpha}_{b} \gg 1.
\ee
Notice that this regime with large $\tilde{\alpha}_b$ is qualitatively
the same as in the NJL model with the cutoff of order $|eB|$ and the
dimensional coupling constant $G \simeq \alpha_{b} /M_{\gamma}^{2} 
= \pi/(2N|eB|)$.

Therefore our analysis shows that there are two opposite regimes of
dynamics of spontaneous symmetry breaking in QED in a magnetic field at
large number of flavors. The first of them, which develops for
$\tilde{\alpha}_b\ll 1$, is essentially the same as the weakly coupled
regime with a small number of fermion flavors $N$. The other limiting case
appears when $\tilde{\alpha}_b\gtrsim 1$, and it is characterized by
pairing dynamics governed by an almost local interaction. In terms of the
number of fermion flavors, these two regimes occur for $N\ll 1/\alpha_b$
and for $N\gtrsim 1/\alpha_b$, respectively.

\section{Conclusion}

QED in an external magnetic field yields an example of a rich dynamics. It
is important that this dynamics can be taken under control both for a weak
coupling constant $\tilde{\alpha}_b$ with $N$ of order one and for an
arbitrary value of $\tilde{\alpha}_b$ when $N \gg 1$. In accordance with
the general analysis of Refs.~\cite{GMS1,GMS2}, the phenomenon 
of the magnetic catalysis in QED is
universal, although its dynamics varies dramatically with 
increasing $N$.

In this paper we did not discuss the dynamical regime with a strong
coupling constant $\tilde{\alpha}_{b}$ and $N$ of order one (genuine
strong coupling regime). Although in this case the dynamics does not admit
a controllable approximation, one should expect that spontaneous chiral
symmetry breaking in this regime takes place even without an external
magnetic field \cite{FGMS}. An external magnetic field should presumably
enhance the value of the dynamical mass for fermions, as it happens for 
example in the supercritical phase of the NJL model [see Ref.~\cite{GMS2}
and the second paper in Ref.~\cite{QED1}].

\begin{acknowledgments}

The work of V.P.G. was supported by the SCOPES-projects 7UKPJ062150.00/1
and 7 IP 062607 of the Swiss NSF and by the Grant No. PHY-0070986 of NSF
(USA). V.A.M. is grateful for support from the Natural Sciences and
Engineering Research Council of Canada. The work of I.A.S. was partially
supported by Gesellschaft f\"{u}r Schwerionenforschung (GSI) and by
Bundesministerium f\"{u}r Bildung und Forschung (BMBF).

\end{acknowledgments}

\end{document}